\documentclass[letterpaper, 10 pt, conference]{ieeeconf}  

\IEEEoverridecommandlockouts                              

\usepackage{graphicx} 
\usepackage{textcomp}

\title{\LARGE \bf
Spatial Summation of Localized Pressure for Haptic Sensory Prostheses
}

\author{Sreela Kodali$^{1}$, Cihualpilli Camino Cruz$^{1}$, Thomas C. Bulea$^{2}$, Kevin S. Rao$^{2}$ Diana Bharucha-Goebel$^{2}$, \\ Alexander T. Chesler$^{2}$, Carsten G. Bönnemann$^{2}$, and Allison M. Okamura$^{1}$
\thanks{
This work was supported in part by a grant from the National Science Foundation award 2301355 and the intramural research program of the National Institutes of Health (NIH) including the NIH Clinical Center, NCCIH, and NINDS. $^{1}$The authors are with Stanford University, Stanford, CA 94305, USA
{\tt\small \{kodali, aokamura\}@stanford.edu}. $^{2}$The authors are with the National Institutes of Health, Bethesda, MD 20892, USA {\tt\small \{thomas.bulea, diana.bharucha-goebel, kevin.rao, alexander.chesler, carsten.bonnemann\}@nih.gov}}%
}

\begin{document}

\maketitle
\thispagestyle{empty}
\pagestyle{empty}

\begin{abstract}

A host of medical conditions, including amputations, diabetes, stroke, and genetic disease, result in loss of touch sensation. Because most types of sensory loss have no pharmacological treatment or rehabilitative therapy, we propose a haptic sensory prosthesis that provides substitutive feedback. The wrist and forearm are compelling locations for feedback due to available skin area and not occluding the hands, but have reduced mechanoreceptor density compared to the fingertips. Focusing on localized pressure as the feedback modality, we hypothesize that we can improve on prior devices by invoking a wider range of stimulus intensity using multiple points of pressure to evoke spatial summation, which is the cumulative perceptual experience from multiple points of stimuli. We conducted a preliminary perceptual test to investigate this idea and found that just noticeable difference is reduced with two points of pressure compared to one, motivating future work using spatial summation in sensory prostheses. \end{abstract}

\section{Introduction}

The human sense of touch plays a crucial role in our perceptual experiences and enabling motor tasks; diseases and injuries that result in loss of touch sensation can be debilitating. 
A rare and intriguing circumstance is PIEZO2 LOF (Piezo Type Mechanosensitive Ion Channel Component 2 Loss of Function), resulting in complete congenital absence of proprioception, no vibration sense, and impaired sense of light touch -- although deep pressure sensation is preserved. 
For most types of sensory loss, there is no pharmacological treatment or rehabilitative therapy, so sensory substitution is the only option. Visual and auditory channels are overloaded or not well matched to missing touch, motivating a “sensory prosthesis” that provides substitutive haptic feedback to an alternative location on the body. In patients with PIEZO2 LOF, deep pressure is a viable form of feedback~\cite{szczot2018piezo2, typeA_afferents}.

For a haptic device to be used as a sensory prosthesis in daily life, we cannot occlude the hands. The wrist and forearm are compelling locations for feedback due to available skin area and straightforward mounting of devices \cite{ASR_MarciaOMalley}. 
We hypothesize we can improve on our prior arm-mounted haptic device \cite{kodaliWHC23} by invoking a larger range of pressures with spatial summation; to do so, we must first quantify and understand this phenomenon. Spatial summation, a decrease in discrimination threshold with an increase in stimulation area, is a well-studied principle in psychophysics that describes the cumulative perceptual experience from multiple points of stimuli \cite{Greenspan1997-no}. 
Limited work on spatial summation in pressure and deep tissue stimulation motivates this perceptual study \cite{HongTan1994, Greenspan1997-no, CuffAlgometry}. Insight on spatial summation will be used to improve range and resolution of future haptic devices.

\begin{figure}[t]
\begin{center}
\includegraphics[width=\columnwidth]{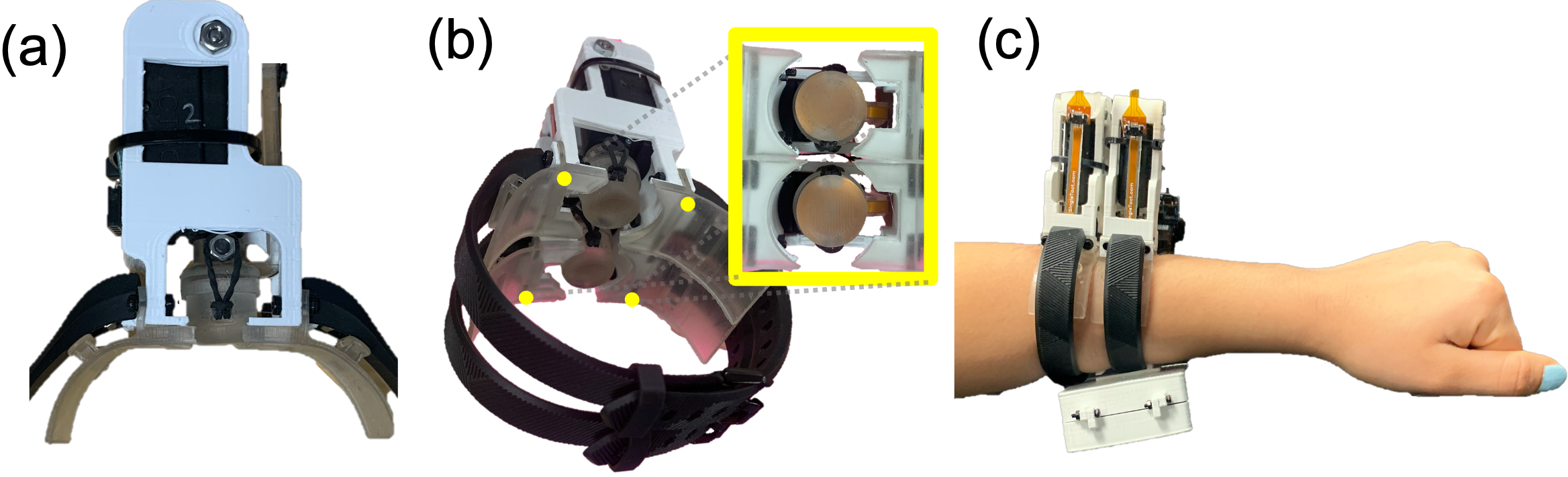}
\end{center}
\vspace{-4mm}
\caption{The wearable device is comprised of modular pressure stimulators, a linear actuator in a rigid enclosure (a). Two  pressure stimulators can be adjacent so their tactors are only 6 mm apart (b) and can be worn with its supporting electronics on the forearm (c).}
\vspace{-5mm}
\label{fig:multipoint}
\end{figure}

\section{Device Design}
We created a wearable device for experimental purposes that can apply localized pressure at two points on the forearm separately or simultaneously, allowing for many combinations of haptic stimuli. The device consists of modular pressure stimulators and supporting electronics (Fig.~\ref{fig:multipoint}).

The pressure stimulator consists of a position-controlled micro linear actuator with position feedback (Actuonix PQ12-30-12-P) mounted in a plastic enclosure. The actuator shaft connects to a cylindrical tactor (15 mm dia.) and an embedded low-profile capacitive force sensor (SingleTact, dia. 15 mm and force range 0-45 N) that allows for real-time force measurement (Fig.~\ref{fig:multipoint}b). The pressure stimulator fastens to the forearm with a flexible interface and elastomer straps (Fitbit\texttrademark Alta). The enclosure, tactor, and flexible interface were all 3D printed. Users can wear multiple modular stimulators and adjust the distance between them. For our preliminary test, we positioned two adjacent pressure stimulators with edges of the tactors 6 mm apart (Fig.~\ref{fig:multipoint}b).

A microcontroller (Arm Cortex M7, Teensy 4.0) serves as the leader, and linear actuator control (LAC) boards for each actuator (Actuonix LAC board) are peripherals. A custom C library was developed for the device's firmware and runs on the microcontroller on a custom PCB. It can interface with up to four pressure stimulators, control and calibrate them independently, and collect force data. 
Each LAC board runs a local programmable PD controller for its corresponding actuator and moves the actuator to a target position; the parameters for the controllers were calibrated individually. All the electronics are enclosed in 3D printed plastic holders that clip to the straps of the pressure stimulators (Fig.~\ref{fig:multipoint}c).

\section{Preliminary Tests and Results}

\subsection{Just Noticeable Difference: One vs.\ Two Contacts} 

We tested the extent of spatial summation by comparing the just noticeable difference (JND) with one and two pressure stimuli from the wearable device, following the methods of Greenspan et al. \cite{Greenspan1997-no}. Quantifying spatial summation will allow us to harness it to generate a wider range of percepts and improve resolution for a future prosthesis. We conducted tests with a single participant. The  hand and elbow were supported while the wrist and forearm were suspended in air. The participant wore headphones playing music to prevent auditory feedback from the actuators, and they could not see the device during the tests. Because the current device is position-controlled, we used the commanded position as a proxy for a pressure stimulus. As the actuator extends, the tactor presses harder on the arm. 
The commanded position was compared with the real-time force measurements, and we found a similar pattern in the force data. 
Although the actuator command is not exactly the same as applied pressure, we propose that it is sufficient for a preliminary test to assess the device and methods.

We measured the participant's minimum detection threshold and maximum threshold before pain with a single pressure stimuli using a modified version of Allowable Stimulus Range (ASR) methods \cite{ASR_MarciaOMalley}. We then measured JND using the method of limits with staircasing, where stimuli are applied in ascending and descending orders depending on the user’s ability to discriminate between pairs of reference and comparison stimuli. The reference value was the midpoint of the participant's ASR. Stimuli pairs were delivered sequentially in random order, and the user indicated whether the first stimulus was greater than, equal to, or less than second stimulus. A 2-down, 1-up staircase design was used, with step sizes set to a prescribed ratio for convergence \cite{Garcia-Perez1998-hq}. 

\begin{figure}[t]
\begin{center}
\includegraphics[width=8cm]{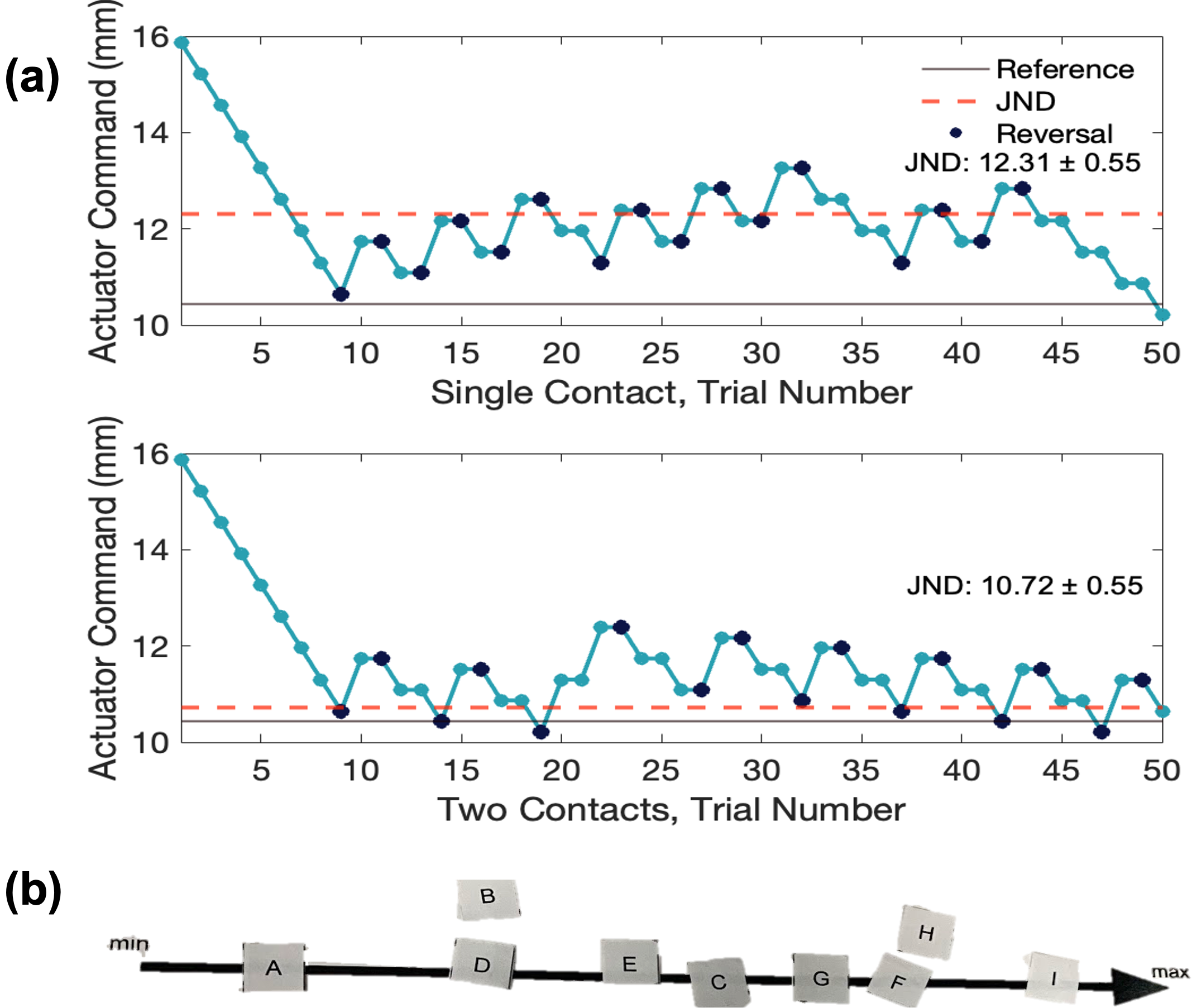}
\end{center}
\vspace{-4mm}
\caption{(a) Results of a 2-down, 1-up staircase for a single stimulus and two stimuli, yielding JNDs of 12.3 ± 0.5 mm and 10.7 ± 0.5 mm, respectively. 
(b) A participant's results from ordering pairs of stimuli. Stimuli A \textit{(MIN, MIN)} and stimuli I \textit{(MAX, MAX)} are positioned at ends of continuum.}
\vspace{-5mm}
\label{fig:staircase}
\end{figure}

The participant went through the staircase method twice: once with a single pressure stimulus and another with two pressure stimuli (Fig.~\ref{fig:staircase}a). 
Both staircases had 16 reversals. JND was computed as the mean of the last 3 reversals \cite{Baths_WHC2023_JND}. The reference value was 10.4 mm, which corresponds to about 4.3 N. For a single pressure stimulus, the JND was 12.3 mm, with a standard deviation of 0.5 mm. For two pressure stimuli, the JND was 10.7 mm with a standard deviation of 0.5 mm. The noticeably lower JND with two pressure stimuli supports the presence of spatial summation and aligns with prior work \cite{HongTan1994}. The force data tracked the actuator command proxy, but the increments were inconsistent due to lack of force control and variable tissue stiffness at different locations in the forearm. Future work will use the embedded force sensor in the device to implement force control. 

\subsection{Ordering Pairs of Stimuli}

The second test asked participants to order, on a continuum, nine pairs of stimuli from the two contact points to observe perceived spatial summation. Each contact point was at one of three states: either minimum pressure, midpoint of ASR, or maximum comfortable pressure, henceforth abbreviated as \textit{MIN}, \textit{MED}, and \textit{MAX}. This resulted in 9 distinct pairs of stimuli assigned letters \textit{A} through \textit{I}: \textit{(MIN, MIN), (MIN, MED), (MIN, MAX), (MED, MIN), (MED, MED), (MED, MAX), (MAX, MIN), (MAX, MED), (MAX, MAX)}. Three participants ordered the stimuli with an older version of the device (Fig.~\ref{fig:staircase}b). The participants chose the same maximum and minimum stimuli and positioned similar stimuli at the 1st and 3rd quartiles in the continuum, indicating that they can rank the intensity of mixed single and multiple-contact stimuli appropriately. This indicates promise for using a combination of individual stimuli and spatial summation to increase the resolution and range of intensity of a haptic sensory prosthesis for the arm.










\bibliographystyle{IEEEtran}
\bibliography{myBib}

\end{document}